\documentstyle[prl,aps,preprint,tighten]{revtex}
\begin{document}
\draft
%\twocolumn[
%\hsize\textwidth\columnwidth\hsize\csname @twocolumnfalse\endcsname
\title{Charge Disorder in Granular Metallic Films}
\author{S.V.Panyukov$^a$ and A.D.Zaikin$^{a,b}$}
\address{$^a$ I.E.Tamm Department of Theoretical Physics,
P.N.Lebedev Physics Institute, Leninskii prospect 53, 117924
Moscow, Russia \\
$^b$ Institut f\"ur Theoretische Festk\"orperphysik, Universit\"at
Karlsruhe, 76128 Karlsruhe, FRG}

\maketitle

\begin{abstract}
It is shown that practically any physically relevant random distribution
of offset charges
destroys the Kosterlitz-Thouless-Berezinskii charge-unbinding phase
transition in two dimensional normal and superconducting granular
arrays and films. The array conductance obeys the Arrenius dependence
on temperature. Offset charge disorder decreases the effective
Coulomb gap of the system and may account for recent experimental
findings in two dimensional arrays of tunnel junctions.
\end{abstract}

\pacs{PACS numbers: 73.40 Gk, 73.40 Rw}
%]
\narrowtext

Charging effects in tunnel junction arrays remain under intensive
investigation during last several years. These effects
become important at sufficiently low temperatures $T \lesssim E_C=e^2/2C$
where $C$ is the characteristic capacitance of a tunnel junction between
metallic islands and/or of such an island itself.
Under this condition tunneling of a
single electron from one island to another is essentially suppressed
due to the Coulomb interaction and the system conductance $G$ shows
an activation behavior $G(T) \propto \exp (-aE_C/T)$, where $a \sim 1$.
At $T=0$ electron  tunneling is blocked completely and the system becomes
insulating.

The above arguments are quite general and apply to the most of
mesoscopic granular arrays. It was pointed out by Mooij {\it et al.}
\cite{M,FS} that in two dimensional (2D) arrays a new interesting
collective effect may take place.
Provided the island capacitance to the ground $C_0$ is small as compared
to the junction capacitance $C$ electrostatic interaction of charges \cite{FN}
in such arrays logarithmically depends on the distance $\Lambda$ between
these charges up to $\Lambda \sim \sqrt{C/C_0}$, $\Lambda$
is measured in the lattice spacing units. Thus in the limit $C_0 \rightarrow 0$
the system represents an example of a 2D Coulomb gas which
exhibits a Kosterlitz-Thouless-Berezinskii (KTB) phase transition \cite{KTB,RG}
at a temperature $T_{KTB}$ of order $E_C$. Physically it implies that
at $T > T_{KTB}$ there is a nonzero concentration of free charges
in the system and its conductance remains finite $G>0$, whereas
at $T <T_{KTB}$ all charges are bound in charge-anticharge pairs and
the system linear conductance drops to zero $G=0$. Thus 2D arrays
may become insulating not only at $T=0$ but also at finite $T < T_{KTB}
\simeq E_C/4\pi$ \cite{M,FS}.

The above conclusion applies to both normal and
superconducting granular arrays. In the latter case an elementary
charge is that of a Cooper pair $2e$ (instead of $e$ for normal
arrays) and therefore $T_{KTB}$ for superconducting arrays is four
times larger than that for normal arrays \cite{M,FS}.

Several experiments \cite{T,D,Zant} were performed to study
charging effects in both normal and superconducting 2D granular
arrays and none of them indicated the presence of a KTB phase
transition for charges. E.g. no specific KTB dependence
$G(T) \sim \exp (-A/\sqrt{T_{KTB}-T})$ in the vicinity of
$T_{KTB} \sim E_C$ has been found. In contrast for a wide
temperature region the array conductance was reported to
follow a purely activation behavior $G(T) \propto \exp (-aE_C/T)$,
where for most of the samples the parameter $a$ varied between
0.23 and 0.27 \cite{M,T,D,Zant}.
These results might look somewhat surprising. Of course in real
systems finite
size as well as the selfcapacitance effects turn a charge-KTB
transition into a crossover. Nevertheless for the sample parameters
\cite{M,T,D,Zant} one estimates $\Lambda \sim 10 \div 20$ and thus
such a crossover could be expected
to be sufficiently sharp to be distinguished from a purely
activation behavior.

An important assumption made in \cite{FS} was that the island charge
is quantized in units of $e$ (or $2e$ for superconducting arrays). In
real experimental situation, however, this assumption is frequently
violated because of the random charges trapped in the substrate. These charges
can polarize the islands and induce noninteger offset charges on them.
Some consequences of this effect were
previously studied numerically \cite{BGFS,MW} and experimentally \cite{C,L}.

In this Letter we argue that randomly distributed
noninteger offset charges have a dramatic impact on the behavior
of 2D granular arrays and that the experimental results \cite{M,T,D,Zant}
acquire a natural physical explanation provided the effect
of offset charges is taken into account. We demonstrate that already
a very weak disorder completely washes out the charge-KTB transition in 2D
arrays leading to an activation-type behavior of the system
conductance $G(T)$. We also show that in the presence of the offset
charges the Coulomb gap for electron tunneling across
a half of junctions in the
array decreases for any given direction of the applied current.
At sufficiently low
temperatures these junctions yield
a dominating contribution to the system conductance and the effective
Coulomb gap becomes smaller than that in the absence of the offset charges.

Let us consider a square tunnel junction array with the junction
capacitance $C$ and assume that the selfcapacitance of the islands
is very small $C_0 \rightarrow 0$. The partition function of this array can
be written in terms of a path integral
(see e.g. \cite{SZ,FS})
\begin{equation}
Z=\sum_{n_x}\exp (\frac{2\pi i n_xQ_x}{e})
\prod_x\int d\phi_{x0}\prod_x\int_{\phi_{x0}}^{\phi_{x0}+4\pi n_x}
 {\cal D}\phi_x\exp \biggl(-\int_0^{\beta} d\tau
\sum_{\langle xx'\rangle }\frac{C}2\biggl(\frac{\dot{\phi}}{2e}\biggr)^2-
S_t[\phi ]\biggr),
\label{Z0}
\end{equation}
where the collective variable $\phi_x(\tau )$ is linked to the voltage
$V_x$ on the island
$x$ by means of a standard relation $\dot{\phi }=2eV_x(\tau )$, $Q_x$ is
the island charge, $\beta =1/T$ and the term $S_t[\phi ]$ describes
electron tunneling between
islands. In a superconducting array the variable
$\phi_x(\tau )$ represents the phase of the superconducting order parameter
of the island whereas in normal arrays it is not more than a formal variable.
The sum in (\ref{Z}) is taken over all integer winding numbers
$n_x$ on each island $x$.

In what follows we shall assume that $quenched$ offset charges $Q_x$ are
randomly
distributed over the system, so that $\overline{Q_x}=0$ and
\begin{equation}
\overline{Q_{{\bf x}}Q_{{\bf x}^{\prime }}}=\frac{e^2}{4\pi^2}
g({\bf x}-{\bf x}^{\prime}).  \label{g}
\end{equation}
Depending on the physical situation different
types of disorder can be considered. E.g. the values $Q_x$ on each
island can be fixed by strong local potentials in the
substrate being completely
independent of each other. Then we have $g(x-x')=g_u\delta (x-x')$ or,
equivalently,
\begin{equation}
g_k=g_u,
\label{local}
\end{equation}
$g_k$ is the Fourier component of $g(x-x')$. Alternatively one can
assume that Coulomb interaction between offset charges on different
islands dominates over the local potentials. In this case we have
\begin{equation}
g_k=4\left[ \sin ^2(k_1/2)+\sin ^2(k_2/2)\right]g_c \equiv 4\Delta
(\bbox{k})g_c,
\label{corr}
\end{equation}
where for the model adopted here \cite{FN1} $g_c=2\pi^2T/E_C$
and the components $k_{1,2}$ of the
wave vector $\bbox{k}$ are normalized by the inversed lattice spacing constant
$1/a$. We believe that practically any physically realistic situation can be
described by a proper combination of (\ref{local}) and (\ref{corr}).

Let us first assume that tunneling between islands is small and disregard
the term $S_t$ in the expression for the partition function (\ref{Z0}).
Then following \cite{FS} we make a shift $\phi_x(\tau) = \phi_{x0}+
4\pi n_x\tau /\beta +\theta_x(\tau )$ and perform Gaussian integration
over the variables $\theta_x (\tau )$ with the boundary conditions
$\theta_x(0)=\theta_x(\beta )=0$. Then we get
\begin{equation}
Z=\sum_{\{n_{{\bf x}}\}}\exp \left[ -
\frac{K}2\sum_{\left\langle {\bf x},{\bf x}%
^{\prime }\right\rangle }\left( n_{{\bf x}}-n_{{\bf x}^{\prime }}\right)
^2+\sum_{{\bf x}}2\pi iQ_{{\bf x}}n_{{\bf x}}/e\right],   \label{Z}
\end{equation}
where we denote $K=2\pi^2T/E_C$. For the sake of definiteness let us
first consider the case of a correlated disorder (\ref{corr}).
In order to average the free energy $F$ over disordered offset charge
configurations we make use of the replica trick and write
\begin{equation}
F=-T\left. \frac d{dm}\overline{Z^m}\right| _{m=0},  \;\;
\overline{Z^m}=\sum\limits_{\{n_{{\bf x}}^\alpha \}}\exp \left[
-\sum\limits_{\alpha =1}^m \frac{K}2\sum
\limits_{\left\langle {\bf x},{\bf x}^{\prime
}\right\rangle }\left( n_{{\bf x}}^\alpha -n_{{\bf x}^{\prime }}^\alpha
\right) ^2-\frac{g_c}2\sum\limits_{\langle {\bf x},{\bf x}^{\prime}\rangle }
\left( \sum\limits_{\alpha =1}^m (n_{{\bf x}}^\alpha
 -n_{x^{\prime}}^{\alpha})\right)
^2\right].   \label{Zm}
\end{equation}

Performing the standard transformation to the field theory
\cite{Villain} with the
aid of the Poisson resummation formula we arrive at the expression
$$
\overline{Z^m}=\int D\varphi \exp \left\{ -\sum_{\left\langle {\bf x},{\bf x}%
^{\prime }\right\rangle }
\left[ \sum_{\alpha =1}^m \frac{K}2\left( \varphi _{{\bf x}%
}^\alpha -\varphi _{{\bf x}^{\prime }}^\alpha \right) ^2+g_c\sum_{\alpha
,\beta =1}^m\left( \varphi _{{\bf x}}^\alpha
 -\varphi _{{\bf x}^{\prime }}^\beta
\right) ^2\right] +\right.
$$
\begin{equation}
\left. \sum\limits_{{\bf x}}
\sum\limits_{\alpha =1}^m 2y\cos \left[ 2\pi \varphi
^\alpha ({\bf x})\right] \right\},  \label{Zm2}
\end{equation}
where $\varphi $ is the $m$-component field. We introduced here the chemical
potential $\ln y$ which unrenormalized value is set equal to zero.

In order to proceed with the renormalization group (RG) analysis we
come to  a continuum limit
$\varphi_x-\varphi_{x^{\prime }} \rightarrow a\nabla \varphi (x)$ and
introduce the minimum length scale $\xi \sim a$. The idea of calculation
is essentially the same as that of \cite{RG}. The generalization concerns
only the effect of disorder.
Expanding the exponent in the expression for $\overline{Z^m}$ in powers of
$y$ and $g$ and successively
integrating out the Fourier components $\varphi_{q'}^{\alpha}$ with
$\bbox{q}-\delta \bbox{q}<\bbox{q'}<\bbox{q}$ one comes to larger and
larger scales $e^l$ and arrives at the RG equations
\begin{equation}
\frac{d\ln y}{dl}=\left( 2-\frac \pi K+\frac{4\pi ^3g}{K^2}\right), \;\;\;
\frac{dK}{dl}=\pi^3 y^2, \;\;\; \frac{dg}{dl}=0.
\label{RG}
\end{equation}
Here we have to put $g=g_c$.

An analogous calculation can be carried out for the uncorrelated
distribution of the offset charges (\ref{local}). In this case
the last term in the square brackets of (\ref{Zm}) has the form
$-(g_u/2)\sum_x(\sum_{\alpha }n_x^{\alpha })^2$.
Repeating the same procedure we again arrive at the RG equations
(\ref{RG}) with $g=g_u$.

It is interesting to point out that the scaling equations (\ref{RG}) have
the same form as those derived earlier in Ref. \cite{RSN} for a 2d XY
model with random phase shifts. However an important physical difference
is that in Ref. \cite{RSN} the case of random $dipols$ (or,
equivalently, $anticorrelated$ charges) was considered whereas here we
are dealing with random offset $charges$ $Q_x$. Apparently at large distances
random charges should be more efficient in screening the
charge-anticharge logarithmic interaction than random
dipoles. And indeed -- although the disorder term in our equations (\ref{RG})
has the same form as in Ref. \cite{RSN} -- the effective disorder strength
in our case turns out to be larger \cite{FN*} than that in the case of random
dipoles \cite{RSN,NK}.

 From the equations (\ref{RG})
it is easy to see that for $g>1/32\pi$ the fugacity $y$ monotonously
increases during the scaling procedure $dy/dK >0$ for all values of $K$.
Therefore the density of free charges in the system remains finite
at any temperature and the KTB phase transition for charges never occurs.
In the case of a correlated disorder (\ref{corr}) the condition
$g_c>1/32\pi$ is equivalent to $T>E_C/64\pi^3 \sim 5 \times 10^{-4} E_C$.
For typical experimental values of $E_C$ \cite{M,T,D,Zant} this inequality
can be violated only at temperatures below 1 mK. For the
uncorrelated disorder (\ref{local})
the condition $g_u >1/32\pi$ includes most of physically important disorder
configurations. E.g. if we assume that offset charges on islands are uniformly
distributed between $-e/2$ and $e/2$ we get the value
$g_u=\pi^2/3 \gg 1/32\pi$ and thus the KTB-ordered phase for charges should
be completely destroyed by
disorder.  Note that this conclusion is consistent with the
numerical results
\cite{BGFS} which demonstrate smearing of a universal jump for
the nonlinear conductance of a 2D array with uncorrelated charge disorder.

For $g < 1/32\pi$ the RG flow (\ref{RG}) becomes nonmonotoneous for not
very large $K_-<K<K_+$ \cite{RSN} (see also fig.1). It was
concluded \cite{RSN} that in this case the
phase diagram of a disordered 2D XY
model shows a reentrant behavior. Recently this
conclusion was criticized by Nattermann {\it et al.} \cite{NK} who
found no reentrant behavior and, making use of the convention about
nonrenormalizability of $T$ (which was not used in \cite{RSN} and here),
argued in favour of the existence
of a KTB phase under the condition equivalent to $g < 1/32\pi$.
We believe that for our physical situation charge disorder should
destroy the KTB phase even for $g < 1/32\pi$. Indeed in our model
(in contrast to \cite{RSN,NK}) the initial (unrenormalized) fugacity $y=y_0$
is not small ($y_0 \sim 1$)
and does not depend on $K$. Integrating the equations
(\ref{RG}) it is
easy to show that for $g\gtrsim (1/8\pi )\exp (-2-2\pi^2 y_0^2)$
the line $y=y_0$ can intersect only the
RG flow lines with $y$ renormalized to larger values (fat solid
lines on fig. 1). Thus even if -- improving the accuracy of the RG analysis
-- we choose somewhat smaller values of $y_0$ (this might only favour
the KTB phase), we still can conclude that no charge-KTB phase transition
occurs down to exponentially small $g$. Therefore there is a
very little (if any) chance to provide experimental conditions for detecting
the KTB phase transition for charges. This in fact was demonstrated in several
experiments \cite{T,D,Zant}.

In order to evaluate the conductance of a 2D granular array the term
$S_t[\phi ]$ (\ref{Z0}) describing electron tunneling between islands
should be taken into account. In the case of normal granular arrays
this term has the form (see e.g. \cite{SZ,FS})
\begin{equation}
S_t[\phi ]= \sum_{\langle x,x^{\prime }\rangle }\int
d\tau \int d\tau '{\alpha _tT^2\over \sin ^2(\pi T(\tau -\tau '))}\cos
 ({\varphi (\tau )-\varphi (\tau ')\over 2}),
\label{St}
\end{equation}
where $\alpha_t=\pi /2e^2R_t$, $R_t$ is the junction tunneling resistance.
In the weak tunneling limit $\alpha_t \ll 1$ one can proceed
perturbatively and expand (\ref{Z}) in powers of $\alpha_t$. The
corresponding analysis is a straightforward generalization of that
for a single junction \cite{SZ}. Keeping only the first order terms
and performing Gaussian integration over the phase variables $\phi_x$
for a given distribution of the offset charges $Q_x$ in the array we obtain
\begin{equation}
F[Q(\bbox{k})]=F_0[Q(\bbox{k})]+\sum_{\langle x,x^{\prime}\rangle }
\int_0^{\beta }d\tau \alpha (\tau ) \exp \biggl[
\sum_k E_C (Q(\bbox{k})/e-1)\tau
\frac{\sin^2(k_1/2)}{\Delta (\bbox{k})}\biggr],
\label{F1}
\end{equation}
where $F_0[Q(\bbox{k})]=\sum_k Q^2(\bbox{k})/2C\Delta (\bbox{k})$
is the free energy of a random distribution of charges $Q(\bbox{k})=
\sum_x Q_x \exp (i\bbox{kx})$ in the array for $\alpha_t=0$. The conductance
of a single junction in the array (shunted by all other junctions) can be
calculated by means of a standard
technique \cite{SZ} which allows to evaluate an imaginary part of the
free energy $F$ by means of an analytic continuation. Proceeding in much
the same way as it has been done in \cite{SZ} for a single junction
we get
\begin{equation}
G[Q(\bbox{k})]=(2R_t)^{-1}\exp \biggl[-\frac{E_C}{2T}\biggl(1- \sum_k
\frac{2Q(\bbox{k})\sin^2(k_1/2)}{e\Delta (\bbox{k})}\biggr)\biggr].
\label{GQ}
\end{equation}
The expression (\ref{GQ}) defines the conductance of a normal tunnel
junction in the array with a fixed distribution of the offset charges
$Q(\bbox{k})$. According to this result in the absence of the
offset charges
$Q(\bbox{k})=0$ the conductance of each junction in the array reads
$G = (2R_t)^{-1}\exp (-E_C/2T)$. As the total conductance $G_{tot}$
of an ordered 2D square array of junctions is equal to
$G_{tot}=2G$ we conclude that the effective Coulomb gap of such an
array is equal to $E_C/2$ \cite{FN2}.

In order to find the average
value $\overline{G}$ for nonzero $Q(\bbox{k})$ one should
integrate (\ref{GQ}) over all disorder configurations. In the case
of a correlated disorder (\ref{corr}) this integration is performed
with the weight factor $\exp (-F_0[Q(\bbox{k})]/T)$. As a result we obtain
\begin{equation}
\overline{G(T)} = \exp [-(1/2-1/\pi )E_C/T].
\label{G}
\end{equation}
Provided the disorder in the distribution of local conductances is not
very large the array conductance $G_{tot}$ can be evaluated within the mean
field approximation which yields  $G_{tot}(T) \sim \overline{G(T)}$ and thus
\begin{equation}
G_{tot}(T) \approx R_t^{-1}\exp (-aE_C/T),
\end{equation}
where for a correlated disorder (\ref{corr}) we have
$a=1/2-1/\pi \simeq 0.2$. This value of $a$ is close to that found
experimentally \cite{M,T,D,Zant}
especially if we take into account the finite size
effect which would make the Coulomb gap somewhat larger than that for an
infinite array.

Thus we came to the conclusion that disorder in the distribution of the offset
charges decreases the Coulomb gap of 2D tunnel junction arrays. This result
has a very transparent physical origin. Indeed in the presence of randomly
distributed offset charges $Q_x$ the effective charge of a half of the tunnel
junctions in the array is positive. Due to that the Coulomb gap for
electron tunneling across such junctions in one direction becomes smaller
than that without an external charge $E_C/2$. At sufficiently low $T$ these
junctions yield a dominant contribution to the array conductance $G_{tot}$ for
the corresponding direction of the
current. For the current of the opposite sign
another half of the junctions with the negative effective charge
would favour electron tunneling and the
Coulomb gap is again smaller than $E_C$.
Hence, in the presence of a charge disorder
one always has $a<1/2$. The particular
value of $a$ may depend on the type of disorder and also on temperature
(the latter is the case e.g. for $g=g_u$ (\ref{local})).

The same type of arguments apply to superconducting arrays. At sufficiently
low temperatures the number of quasiparticles above the superconducting gap
$\Delta_0$ is exponentially small and we get
$G_{tot} \propto \exp [-(aE_C+\Delta_0)/T]$. This result is also in a good
agreement with the experimental findings \cite{T,Zant}.

In conclusion, we demonstrated
that in the presence of quenched randomly distributed
offset charges the charge-KTB transition in 2D granular arrays and films
is completely destroyed already in the small disorder limit. The array
conductance shows an activation-type temperature dependence with the
effective Coulomb gap being smaller than that in the absence of offset
charges.  We believe that our results essentially explain the experimental
data \cite{M,T,D,Zant} obtained for low conductance 2D junction arrays.

We are grateful to R. Fazio and G. Sch\"on for useful discussions and
to S.E. Korshunov for clarifying remarks concerning the paper
\cite{NK}. One of us (A.D.Z.) acknowledges the support by the
Alexander von Humboldt
Stiftung and the hospitality of ISI in Torino where the part of this work
has been completed.

%\begin{figure}
%\caption{The Rubinstein-Schraiman-Nelson RG flow in the limit of small
%disorder ($g<1/32\pi$ for the equations (8)). For sufficiently large
%values of initial fugacity $y_0$ and except for exponentially small $g$
%the trajectories corresponding to a renormalized to zero fugacity $y$ (thin
%solid lines between $K_-$ and $K_+$,
%$K_{\pm}=\pi /4 \pm (\pi^2/16-2\pi^3g)^{1/2}$)
%are irrelevant and $y$ always increases during the scaling procedure (fat
%solid lines).}
%\label{fig.1}
%\end{figure}

\begin{references}
\bibitem{FN} An island can be
charged if one electron tunnels to this island or leaves it.  Therefore
in the case of normal arrays an elementary charge is equal to $\pm e$.
\bibitem{FN1} Here we assume that the Coulomb interaction between offset
charges is confined to a thin layer close to the substrate surface  and
is governed by the capacitances $C$. Other physical choices of $g_c$ (e.g.
one with $g_c$ being independent of temperature) are also possible.
\bibitem{FN*} The disorder terms in the RG equations \cite{RSN} and (\ref{RG})
become identical if we put $g=\sigma /4\pi^2$, where $\sigma$ is the
disorder distribution width for dipoles \cite{RSN}. It means that
a strongly disordered dipole
configuration with $\sigma \sim 1$ effectively corresponds to a still
weak disorder for charges $g \ll 1$ and vice versa.
\bibitem{FN2} This result can be easily understood because for $C_0=0$ and
low $T$ the only elementary excitation in the system is a charge-anticharge
pair created by an electron tunneling event between two adjacent islands.
As the effective junction capacitance in a 2D square array is $2C$ the energy
of this excitation is equal to \cite{T} $e^2/4C=E_C/2$.
\end{references}
\end{document}